# WSe$_2$/WS$_2$ moiré superlattices: a new Hubbard model simulator

Yanhao Tang[1], Lizhong Li[1], Tingxin Li[1], Yang Xu[1], Song Liu[2], Katayun Barmak[3], Kenji Watanabe[4], Takashi Taniguchi[4], Allan H MacDonald[5], Jie Shan[1,6,7]\*, Kin Fai Mak[1,6,7]\*

[1]School of Applied and Engineering Physics, Cornell University, Ithaca, NY, USA
[2]Department of Mechanical Engineering, Columbia University, New York, NY, USA
[3]Department of Applied Physics and Applied Mathematics, Columbia University, New York, NY, USA
[4]National Institute for Materials Science, 1-1 Namiki, 305-0044 Tsukuba, Japan
[5]Department of Physics, University of Texas at Austin, Austin, TX, USA
[6]Laboratory of Atomic and Solid State Physics, Cornell University, Ithaca, NY, USA
[7]Kavli Institute at Cornell for Nanoscale Science, Ithaca, NY, USA
Email: jie.shan@cornell.edu; kinfai.mak@cornell.edu

**The Hubbard model, first formulated by physicist John Hubbard in the 1960s [1], is a simple theoretical model of interacting quantum particles in a lattice. The model is thought to capture the essential physics of high-temperature superconductors, magnetic insulators, and other complex emergent quantum many-body ground states [2-6]. Although the Hubbard model is greatly simplified as a representation of most real materials, it has nevertheless proved difficult to solve accurately except in the one-dimensional case [4,6]. Physical realization of the Hubbard model in two or three dimensions, which can act as quantum simulators [7], therefore have a vital role to play in solving the strong-correlation puzzle. Here we obtain a quantum phase diagram of the two-dimensional triangular lattice Hubbard model by studying angle-aligned WSe$_2$/WS$_2$ bilayers, which form moiré superlattices [8] because of the difference in lattice constant between the two two-dimensional materials. We probe both charge and magnetic properties of the system by measuring the dependence of optical response on out-of-plane magnetic field, and on gate-tuned carrier density. At half filling of the first hole moiré superlattice band, we observe a Mott insulating state with antiferromagnetic Curie-Weiss behavior as expected for a Hubbard model in the strong interaction regime [2,4,6,9-12]. Past half filling, our experiment suggests an antiferromagnetic to paramagnetic quantum phase transition near 0.6 filling. Our results establish a new solid-state platform based on moiré superlattices which can be used to simulate outstanding problems in strong correlation physics that are manifested by triangular lattice Hubbard models.**

The Hubbard model Hamiltonian has only two terms [1-6], a kinetic lattice hopping term (*t*) and an on-site Coulomb repulsion interaction term (*U*). The Hubbard model on a triangular lattice is particularly rich because geometric frustration enhances quantum fluctuations and leads to a multitude of closely competing states [5,6,10-14] that challenge theory [6,14]. The discovery of superconductivity and correlated insulating states in twisted graphene systems has opened a new chapter in strong electronic correlation physics [15-24]. A recent theoretical analysis has shown that moiré superlattices formed by semiconductor transition metal dichacogenide (TMD) heterobilayers provide a realization of the triangular lattice Hubbard model [8]. Unlike graphene systems, in which spin, valley, and layer degeneracies are all present, the spin and layer degeneracies in TMD heterobilayers are lifted by the strong spin-orbit interactions [25] and the layer asymmetry, respectively. When the chemical potential is within the topmost valence band, where there is large spin-orbit splitting [25], holes can be mapped to the single-band Hubbard model with a two-fold valley degree of freedom, which acts as spin. The model parameters are widely tunable ($U \sim 40 - 50$ meV and $t \sim 0 - 10$ meV) by varying the TMDs in the heterobilayer and their twist angle [8]. Here we study the phase diagram of aligned WSe$_2$/WS$_2$ bilayers around half filling in the large $U/t$ limit by performing optical spectroscopy on moiré excitons [26-30] supplemented by transport measurements. Heat extraction via the phonon bath and simple gating procedures allow the Hubbard model phase diagram to be explored over a wider range of density and temperature than what is achievable in cold-atoms in optical lattices [31].

There is a ~ 4% lattice mismatch between WSe$_2$ and WS$_2$ [32], which produces a triangular moiré superlattice with a periodicity close to 8 nm for an angle-aligned bilayer (Fig. 1a). Both AA (0°) and AB (60°) stackings have been created. (See Methods and Extended Data in Fig. 1 for details on the angle-aligned transfer, and on the determination of the crystal orientations by angle-resolved second harmonic generation.) To introduce electrons or holes into the moiré superlattice, we sandwich the bilayer between two symmetric gates (Fig. 1c). Each is made of a hexagonal boron nitride (hBN) gate dielectric of ~ 20 nm thickness and a few-layer graphene gate electrode. Applying equal voltages ($V_g$) on both gates varies the carrier density in the bilayer while keeping the vertical electric field unchanged. The AA and AB stackings (total 7 samples studied) show similar results. We present results from an AB-stacked sample below, and results from an AA-stacked sample in Extended Data Fig. 2. Unless otherwise specified, all measurements were performed at 1.6 K.

WSe$_2$/WS$_2$ bilayers are known to have type-II band alignment [8, 28, 33] (Fig. 1b). This is fully consistent with our results for the gate-dependent reflection contrast spectrum of a misaligned bilayer (near 20° twist angle) and an aligned bilayer (near 60° twist angle). In the misaligned bilayer (Fig. 1d), the two most prominent features (reflection contrast dips) are the neutral A exciton resonance of WSe$_2$ and WS$_2$. For gate voltages between the two dashed lines, both A excitons are present, indicating that the Fermi level is inside the energy gap of the bilayer. Positive $V_g$ introduces electrons into the WS$_2$ layer, and negative $V_g$ introduces holes into the WSe$_2$ layer. This is demonstrated by the disappearance of the neutral A exciton and emergence of charged excitons in the corresponding layer, accompanied by the nearly unaffected neutral A exciton in the opposing layer.

In the aligned bilayer (Fig. 1e), multiple peaks (I, II, III) emerge around the original WSe$_2$ and WS$_2$ A exciton resonances when the Fermi level is inside the energy gap of the bilayer ($V_g$ between the two dashed lines). This agrees with moiré exciton properties observed in previous works [26-28]. Multiple flat exciton minibands are generated in each layer by the periodic moiré potential that is much stronger than the exciton kinetic energy. The moiré excitons also exhibit distinct gate dependences from that of the A exciton in the misaligned bilayer. In particular, quasiperiodic modulations are observed in the reflection contrast, which are correlated with half-integer moiré band fillings, as we discuss below. We focus on the effect of filling the first hole moiré band of $WSe_2$ (boxed region), the Hamiltonian of which has been shown to map to the single-band Hubbard model on a triangular lattice [8] when interactions are taken into account.

Figure 2a is the enlarged boxed region. We convert gate voltage to hole density $n$ using the gate capacitance (see Methods for details). The valence band edge was determined as in previous work [34] from the fall edge of the moiré exciton reflection contrast (Fig. 2a, b). The moiré band filling factor $n/n_0$ was calculated using a moiré density $n_0/2 = 2.1 \times 10^{12}$ $cm^{-2}$ (corresponding to a 1° twist angle). The uncertainty in $n/n_0$ is estimated to be about 10%, which is dominated by the uncertainty in twist angle (see angle-resolved second harmonic generation in Methods). The three horizontal dashed lines from top to bottom mark the band edge, half filling (one hole per moiré period), and full filling, respectively.

Figure 2b shows the absolute amplitude of the peak reflection contrast of the lowest-energy exciton (I) as a function of filling factor. Prominent peaks on a smooth background (that decreases with $n/n_0$) are noted at half and full fillings. They are correlated with peaks in the two-point resistance of the $WSe_2$ layer (Fig. 2c). The resistance at low carrier densities is dominated by the large contact resistance, which precludes reliable four-point measurements of the sample resistance. Nevertheless, the substantially stronger increase in resistance with decreasing temperature at these filings than away from them (e.g. at 0.8 filling) supports an insulating state at half and full fillings (inset, Fig. 2c).

While the insulating behavior at full filling of the first moiré band is expected for an isolated band even in the absence of interactions, the insulating behavior at half filling is of a different origin. In the limit of large $U/t$, the strong on-site repulsion suppresses double occupancy of the moiré lattice. Each lattice site, located at the maximum of the effective valence band potential energy, is occupied by one hole, giving rise to a Mott insulator with a gap energy ~ $U$ [2, 3, 8]. We find that the insulating behavior persists up to 150 K (Fig. 2c), providing a rough estimate of 10 – 20 meV for $U$. This value is consistent with the predicted values [8] as mentioned. The large moiré exciton reflection contrast in the insulating states is intriguing and likely arises from reduced free carrier screening of the electron-hole interactions in the insulating states.

The ground and low-energy excited states of the Mott state are determined by interactions between the locked spin/valley degrees of freedom that survive charge localization in the insulator. To gain insight into the magnetic interactions between the localized spins on each triangular lattice site, we measure the helicity-resolved reflection contrast spectrum of the lowest-energy moiré exciton as a function of out-of-plane magnetic field $H$ (Fig. 3a). The out-of-plane field is known to break the valley degeneracy, and each handedness of light couple only to one of the two valleys in this system [35]. In the absence of field, the helicity-resolved exciton resonances are degenerate. Under a finite field, they acquire an energy splitting $E_Z$ and have different reflection amplitudes. We analyze the field dependence of $E_Z$ in Fig. 3b, and the spectrally integrated reflection difference in Extended Data Fig. 3. The two quantities have very similar dependence on external field and we assume that both are proportional to the spin/valley polarization. (See Methods for details on helicity-resolved reflection contrast measurements and the determination of $E_Z$.) At half filling, $E_Z$ increases linearly with field at small fields, and saturates to ~ 9 meV above ~ 1 Tesla. The slope at the origin gives a giant exciton valley g-factor ~ 270 ( $g \equiv \lim_{H \to 0} \frac{E_Z}{\mu_B \mu_0 H}$ with $\mu_B$ and $\mu_0$ denoting the Bohr magneton and the vacuum permeability, respectively). The behavior is diametrically opposite at zero filling: There we find $g_0 \approx -4$ and no saturation up to 8 T (Extended Data Fig. 4). The latter behavior is in excellent agreement with the reported exciton valley Zeeman effect in pristine monolayer $WSe_2$ [35].

Moiré excitons in the presence of one hole per moiré site, with or without a magnetic field, are complex many-body excitations that deserve systematic theoretical study. Here we compare our experiment with a simple phenomenological model based on a mean-field interaction between localized excitons and holes in the moiré superlattice. In addition to the applied field $H$, the excitons 'sense' the magnetic moment of localized holes via a molecular field $\lambda_X M$ that is proportional to the sample magnetization $M$ ($\lambda_X$ is a coupling constant). The exciton Zeeman splitting thus consists of two contributions $E_Z = g_0 \mu_B \mu_0 (H + \lambda_X M)$. From our experiment, the second contribution dominates at low temperatures. The field dependence of $E_Z$ is well described by the Brillouin function (solid line, Fig. 3b) that is commonly used to describe the magnetization of localized spin systems [36]. The first contribution (bare exciton response) becomes visible after $M$ saturates at high fields (Extended Data Fig. 4).

The sensitivity of the moiré excitons to the local magnetic moments provides us with an opportunity to examine the temperature dependent magnetic susceptibility $\chi$ (= $M/H$ in the small-field limit), which can be extracted from the measured exciton g-factor using $\lambda_X \chi = \frac{g - g_0}{g_0}$. Figure 3c shows ($g - g_0$) up to 50 K (we limit our analysis to the temperature range where exciton thermal broadening is negligible). We find that $\chi$ decreases by nearly two orders of magnitude when temperature $T$ increases from 1.6 K to 50 K. The susceptibility $\chi(T)$ follows a Curie-Weiss law [$\chi^{-1} \propto (T - \theta)$] for $T \gtrsim 4$ K, with a Weiss constant $\theta \approx -0.6 \pm 0.2$ K (red lines in Fig. 3c, d). The observed Curie-Weiss behavior at high temperatures ($T \gg |\theta|$) with a negative $\theta$, which implies antiferromagnetic (AF) interactions between localized moments, is consistent with the triangular lattice Hubbard model in the large $U/t$ limit [6, 8, 10-12, 14]. In this limit the Hubbard model can be mapped to a valley pseudospin Heisenberg model with AF superexchange coupling $J \sim -\frac{t^2}{U}$ between moments localized at neighboring moiré sites [8]. The value of $\theta$ provides an estimate for $J \sim -0.05$ meV, which is compatible with the predicted $U$ and $t$ values for TMD heterobilayers [8]. The value also agrees well with estimate obtained by assuming that the valley pseudospins form a three-sublattice 120° non-collinear state (see Methods for details). Given geometric frustration in triangular lattices and the 2D nature of the system, the magnetic ordering temperature is therefore expected to lie well below the range accessible in our study.

Equipped with this understanding of the half-filling case, we extend the analysis to other fillings. Figure 4a shows the field dependence of $E_Z$ away from half filling. $E_Z$ saturates at a similar value at large fields for all fillings, implying that full valley polarization is always achieved (and $\lambda_X$ is weakly filling dependent). (There is also the suggestion of a metamagnetic transition below half filling.) The saturation field decreases monotonically with increasing filling for fillings smaller than about 0.6. Figure 4c summarizes the filling dependence of the exciton g-factor at 1.6 K ($<< t$). The low-temperature g-factor takes the bare exciton value $g_0 \approx -4$ near zero and full fillings (Extended Data Fig. 4), and increases by nearly two orders of magnitude towards the middle, reaching a plateau (~ 300) between 0.5 and 0.7 filling. Increasing the temperature rapidly quenches the steep increase of the g-factor (Extended Data Fig. 5), which is suggestive of a divergence in the zero temperature limit. The divergence is extrapolated to occur near 0.6 filling (vertical dashed line in Fig. 4c). The divergent g-factor implies a divergent magnetic susceptibility.

Figure 4b compares the temperature dependence of $\chi^{-1} \propto |\frac{g_0}{g-g_0}|$ with Curie-Weiss law at various fillings. For all cases the data follows the Curie-Weiss law at high temperatures. The best-fit Weiss constant $\theta$ (Fig. 4c) is negative below 0.6 filling (suggesting AF interactions or Fermi pressure of the moiré band holes), but stays near zero above 0.6 filling suggesting a paramagnetic (PM) phase or weak ferromagnetic (FM) interactions ($\theta > 0$), which is inconclusive given the experimental uncertainty in $\theta$. Far away from half filling $\theta$ cannot be accurately measured, but is expected to drop to zero in the absence of carriers because pristine WSe$_2$ is nonmagnetic.

The divergent magnetic susceptibility together with the continuously vanishing $\theta$ near 0.6 filling suggests an AF-PM quantum phase transition, which has also been speculated by a recent numerical simulation of the triangular lattice Hubbard model [12]. (We however cannot exclude the possibility of a crossover from effective AF to weak FM interactions due to experimental uncertainty.) Our data seems to support the presence of a quantum critical point (QCP) near 0.6 filling. Figure 4d further shows the scaling analysis of $E_Z$ (excluding the bare exciton contribution) as a function of $H/T$. The dependences for temperature spanning more than a decade collapse to one curve, indicating that temperature is the only relevant energy scale, where a nearly perfect Curie dependence $\chi^{-1} \propto T$ is obeyed. Finally, we note that the asymmetric phase diagram around half filling is correlated with the asymmetric flatband of TMD heterobilayers, which has a van Hove singularity in the density of states at ¾ filling [8]. Whether or not superconductivity emerges in WSe$_2$/WS$_2$ bilayers near the QCP deserves further investigation by low temperature transport measurements, which could also shed additional light on the nature of the ground states away from the QCP.

**Methods**
**Sample preparation and device fabrication**

The dual-gated WSe$_2$/WS$_2$ bilayer devices were built from exfoliated van der Waals materials using a layer-by-layer dry transfer method [37]. All atomically thin flakes were exfoliated from bulk crystals onto silicon substrates. Monolayer $WS_2$ and $WSe_2$ flakes were identified by their optical contrast. The crystal orientations of $WS_2$ and $WSe_2$ were determined by optical second harmonic generation (SHG) as described below. Hexagonal boron nitride (hBN) flakes with a thickness of about 20 nm were used as the gate dielectrics. Few-layer graphite or $TaSe_2$ was used as gate electrodes. Devices for optical measurements used few-layer graphite as a contact electrode. Entire devices were therefore assembled and released onto Si substrates with prepatterned Au electrodes in a single step. Devices for transport measurements were built in three steps. First, a bottom gate made of graphite and hBN was prepared on a silicon substrate. Next, Pt electrodes were evaporated onto the bottom gate using electron beam lithography and evaporation. Finally, the rest of the device was assembled and released onto the Pt electrodes. Alignment between the $WS_2$ and $WSe_2$ crystals was achieved using a rotation stage with a precision of about 1°. (Because the two layers have different lattice constants and because there is no magic angle property, the sensitivity of device properties to twist angle is much weaker than in graphene multilayers.) The stamp employed to pick up flakes from substrates was made of a polycarbonate (PC) layer on polypropylene-carbonate-coated polydimethylsiloxane (PDMS). van der Waals materials were released from the stamp to the substrate at about 180℃. The PC residue on the devices was removed by dissolving it in chloroform, followed by a rinse in isopropyl alcohol. Sample transfer was performed in a nitrogen-filled glovebox for better interface quality. Extended Data Figure 6 is an optical micrograph of a typical device.

**Angle-resolved optical second harmonic generation (SHG)**

Angle-resolved SHG was employed to determine the crystal orientations of the TMDs (Extended Data Fig. 1). During the measurement, samples were kept in vacuum with pressure below 0.1 torr to avoid potential optical damage. The output of a Ti:sapphire oscillator (Spectra Physics, Tsunami) with a peak photon energy of 1.55 eV, a repetition rate of 80 MHz, and a pulse duration of 100 fs was used as the excitation source. The excitation beam of below 2 mW was focused onto the TMD flakes under normal incidence through a microscope objective with a numerical aperture (N.A.) of 0.6. The beam diameter was about 2 $\mu$m on the samples. The second-harmonic signal was collected by the same objective in reflection geometry and detected by a spectrometer coupled with a liquid-nitrogen cooled CCD camera. A short-pass filter was used in the detection path to cut off the fundamental light. The excitation was linearly polarized. An achromatic half-wave plate was used to control the excitation polarization with respect to the sample orientation. The reflected second-harmonic beam was passed through the same half-wave plate and an analyzer that was kept perpendicular to the excitation light polarization. The SHG from monolayer TMDs follows a six-fold pattern. The peak corresponds to the zigzag direction of the crystal [38]. In bilayers with AA stacking the SHG is much enhanced, and in bilayers with AB stacking it is largely canceled. To determine the twist angle in bilayers quantitatively, we measured the angle-resolved SHG in the non-overlapping region of the bilayer for both $WS_2$ and $WSe_2$, determined their orientations, and extracted the twist angle assuming that strain and other distortions in the overlapped region are negligible. For the device shown in the main text, the twist angle in the bilayer was determined to be $0.4 \pm 0.2°$ by the angle-resolved SHG measurement (Extended Data Fig. 1). Using lattice constants of 0.328 nm and 0.315 nm for $WSe_2$ and $WS_2$, respectively,[32] we obtain a moiré superlattice constant of $7.8 \pm 0.1$ nm. This gives a moiré density of $n_0/2 = (1.9 \pm 0.1) \times 10^{12}$ cm$^{-2}$, which has a 10 % error compared to the calculated value in Fig. 2, presumably due to unaccounted strain and other distortions in the overlapped region.

**Reflection contrast spectroscopy**

Devices were mounted in a close-cycle cryostat (Attocube, Attodry 2100) for magneto-optical measurements down to 1.65 K. A halogen lamp was used as a white light source for the reflection measurements. The output of the lamp was collected by a single-mode fiber, collimated by a 10× objective, and then focused onto the sample by an objective with N.A. of 0.8. The beam diameter on the sample was about 1 $\mu m$ and the beam power is below 1 nW. The reflected light was collected by the same objective and detected by a spectrometer coupled with a liquid-nitrogen cooled CCD camera. To probe the lowest-energy moiré exciton in WSe$_2$ as a function of doping density, a long-pass filter was used in the excitation path to cut off light below 700 nm to avoid potential photo-doping effects by short-wavelength illumination. A combination of a linear polarizer and an achromatic quarter-wave plate was used to generate circularly polarized white light. The reflectance contrast spectrum $R$ was obtained by comparing the reflected light intensity from the sample ($I'$) to that of a featureless spectrum ($I$) from the sample in the highly hole-doped regime or from the substrate. The reflectance contrast was extracted as $R \equiv (I' - I)/I$.

**Transport measurements**

The electrical measurements were carried out in a close-cycle cryostat (Oxford, TeslatronPT) down to 1.5 K. An ac bias of 10-20 mV at 17 Hz was applied between the source and drain, and the source-drain current was measured with a lock-in amplifier.

**Calibration of doping density**

To calibrate the hole density in WS$_2$/WSe$_2$ bilayers from the gate voltage, we first determine the valence band edges. The fall edge (90%) of the peak reflectance of a neutral A exciton or moiré exciton signifies hole doping into WSe$_2$ (Fig. 1d, e and 2a, b) [34]. Note that the valence band edge determined in this way produces a zero-to-half filling density almost identical to the half-to-full filling density (Fig. 2a, b), confirming the reliability of this procedure. The hole density was obtained from the parallel plate capacitor model $n = \frac{\varepsilon\varepsilon_0 \Delta V_g}{d_T} + \frac{\varepsilon\varepsilon_0 \Delta V_g}{d_B}$. Here $d_T = 21.3 \pm 0.3$ nm and $d_B = 21.4 \pm 0.4$ nm are the thicknesses of the top and bottom hBN dielectrics, respectively, determined from the atomic force microscopy (AFM) measurement; $\Delta V_g$ is the applied gate voltage relative to the location of the valence band edge; $\varepsilon_0$ is the vacuum permittivity; and $\varepsilon \approx 3$ is the dielectric constant of hBN, which was independently measured by the Hall effect measurements and is consistent with reported values [39].

**Determination of Zeeman splitting**

The exciton Zeeman splitting was determined as the difference between the spectrally averaged exciton peak positions for the left- and right-handed reflection contrast channels. The spectrally averaged exciton peak position for each channel is obtained by a weighted sum of photon energy with the exciton reflection peak as a distribution function. The weighted sum was carried over a spectral range, in which the exciton reflection contrast is over 90% of the peak reflection contrast. The result is insensitive to this choice down to 70% of the peak reflection contrast. The analysis based on exciton Zeeman splitting is also in good agreement with that based on spectrally integrated magnetic circular dichroism (MCD) $\frac{\int dE |R_+ - R_-|}{\int dE |R_+ + R_-|}$, as shown in Extended Data Fig. 3. Here $R_+$ and $R_-$ are the reflection contrast spectra for left- and right-handed light, respectively, and we have ignored the contribution of the imaginary part of the optical conductivity in $R$. The integration is carried over a spectral range covering the exciton peak. For small exciton splitting and a reflection contrast spectrum consisting of a well-defined exciton peak, we can write down $\frac{\int dE |R_+ - R_-|}{\int dE |R_+ + R_-|} \approx E_Z \frac{\int dE \left|\frac{dR}{dE}\right|}{2 \int dE |R|} \propto E_Z$. The exciton Zeeman splitting is therefore a good measure of the spectrally integrated MCD.

**Estimate of effective AF interactions at half filling**

The superexchange $J$ can also be estimated by assuming that the valley pseudospins form a three-sublattice 120° non-collinear ground state that is canted by the out-of-plane magnetic field. The fractional valley polarization is given by $f = \frac{E_Z}{E_Z^{sat}} = \frac{\Delta}{9|J|}$, where $\Delta$ is the hole Zeeman splitting in the Heisenberg Hamiltonian and $E_Z^{sat}$ is the saturated value of the exciton Zeeman splitting. Saturation (i.e. full valley polarization $f = 1$), which occurs at ~ 1 T (Fig. 3b), is expected when the hole Zeeman energy overcomes the superexchange coupling. Given a hole g-factor ≈ 8, this interpretation implies $\Delta$ ~ 0.48 meV under saturation and $|J|$ ~ 0.05 meV.

**Data availability**

The data that support the plots within this paper, and other findings of this study, are available from the corresponding authors upon reasonable request.

**Figures**

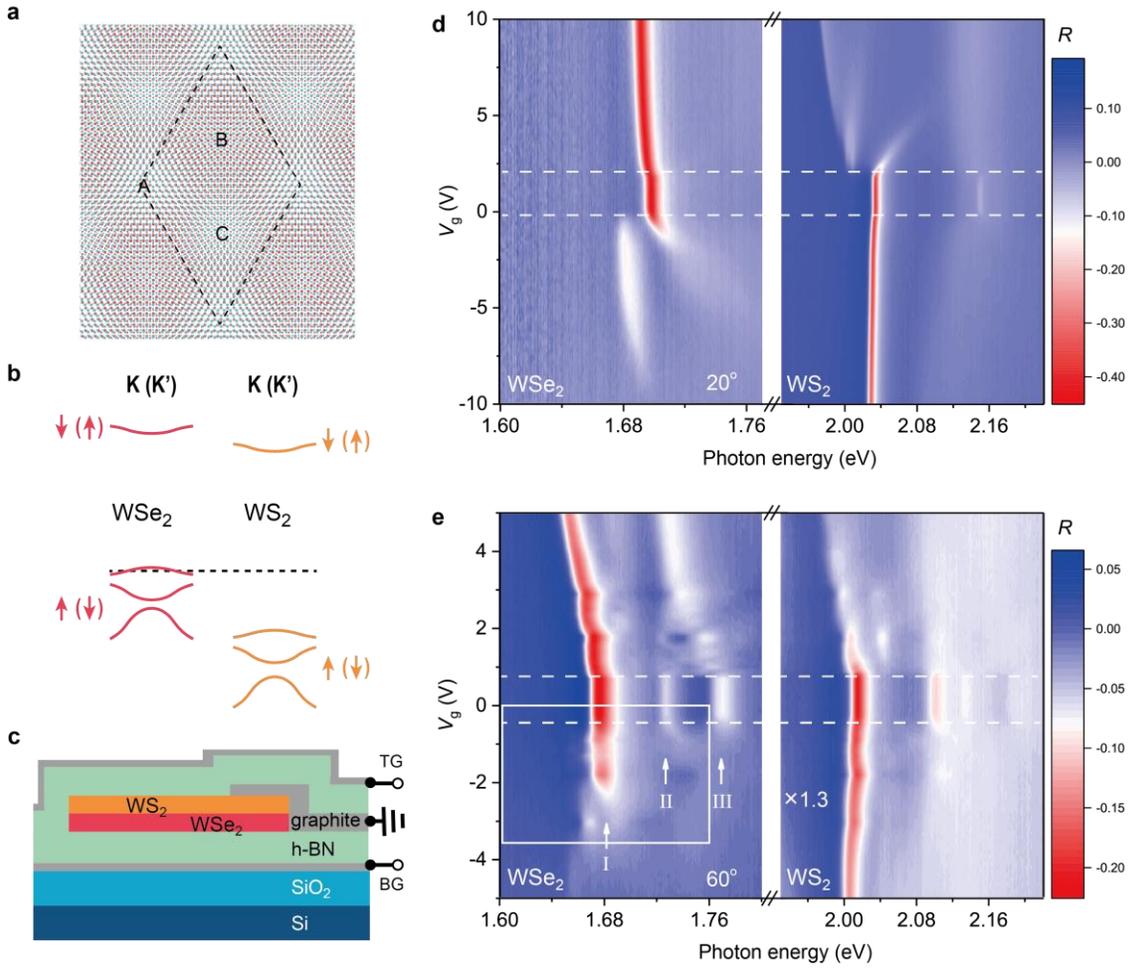

**Figure 1 | Type-II band alignment in WSe$_2$/WS$_2$ bilayers. a,** Illustration of the moiré superlattice formed by an AB-stacked WSe$_2$/WS$_2$ bilayer. The dashed lines outline the moiré unit cell. A, B and C denote three different characteristic moiré sites. **b,** Schematic type-II band structure of the bilayer. Red and orange curves denote bands from WSe$_2$ and WS$_2$, respectively. Arrows denote spin direction. Dashed line shows the Fermi level, which is located in this study near the top most moiré valence band in WSe$_2$. **c,** Schematic geometry of a dual-gate device. **d**, **e**, Contour plot of the gate voltage dependent reflection contrast (*R*) spectrum of WSe$_2$/WS$_2$ bilayers with ~ 20° (**d**) and ~ 60° (**e**) twist angles. Gate voltage $V_g$ is applied only on the top gate in **d** and on both gates in **e**. The two white dashed lines bound the interval of gate voltage over which the Fermi level is inside the energy gap of the type-II band structure. I, II and III in **e** denote the intralayer moiré excitons of WSe$_2$. The rectangular box encloses the lowest-energy moiré exciton over the range of gate voltage where the Fermi level is swept through the top most moiré valence band.

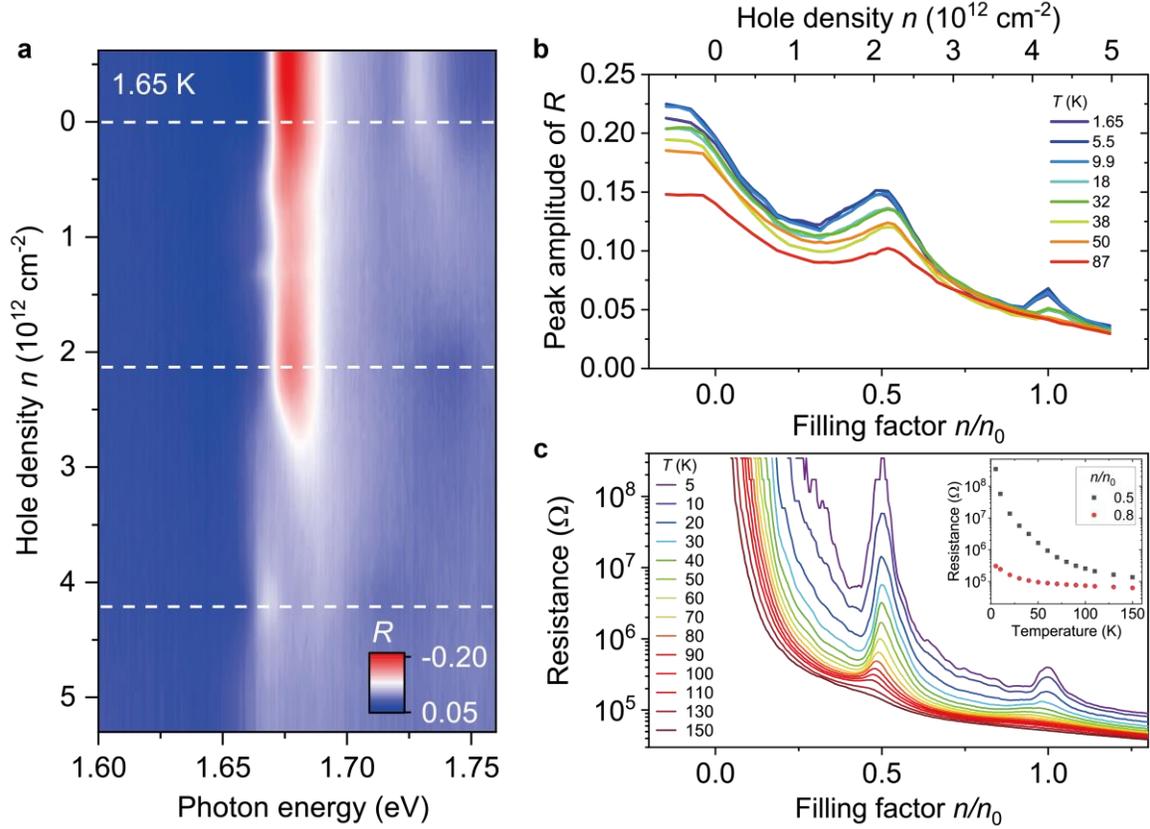

**Figure 2 | Mott insulating state at half filling. a**, Contour plot of doping-dependent reflection contrast spectrum corresponding to the rectangular box in Fig. 1e. The three dashed lines respectively mark, from top to bottom, zero, half and full filling of the first moiré valence band. **b**, Reflection peak amplitudes at the lowest-energy exciton resonance as a function of filling (bottom axis) or hole density (top axis) at different temperatures. **c**, Corresponding two-terminal resistances as a function of filling at different temperatures. The inset shows the temperature-dependent resistance for $n/n_0 = 0.5$ (black) and 0.8 (red).

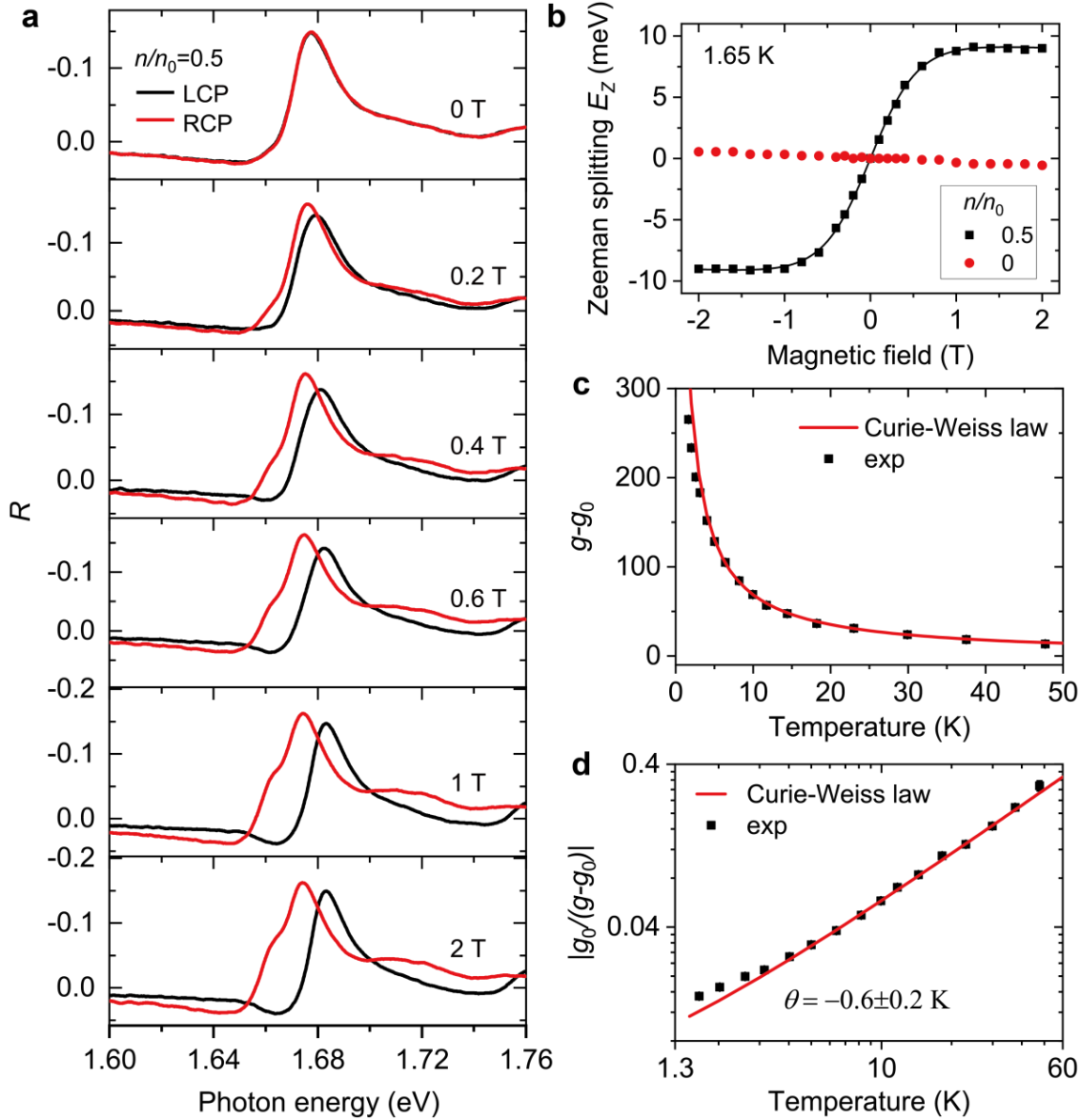

**Figure 3 | Magnetic interactions at half filling. a**, Left- (black) and right-handed (red) reflection contrast spectra under an out-of-plane magnetic field ranging from 0 to 2 T. **b**, Exciton Zeeman splitting ($E_Z$) as a function of magnetic field for half (black squares) and zero filling (red circles). The solid curve is a Brillouin function fit. **c**, **d**, Temperature-dependent $(g - g_0)$ (**c**) and $\left|\frac{g_0}{g-g_0}\right|$ (**d**). The red curves are Curie-Weiss fits from 6.4 K to 48 K, revealing a Weiss constant of $-0.6 \pm 0.2$ K.

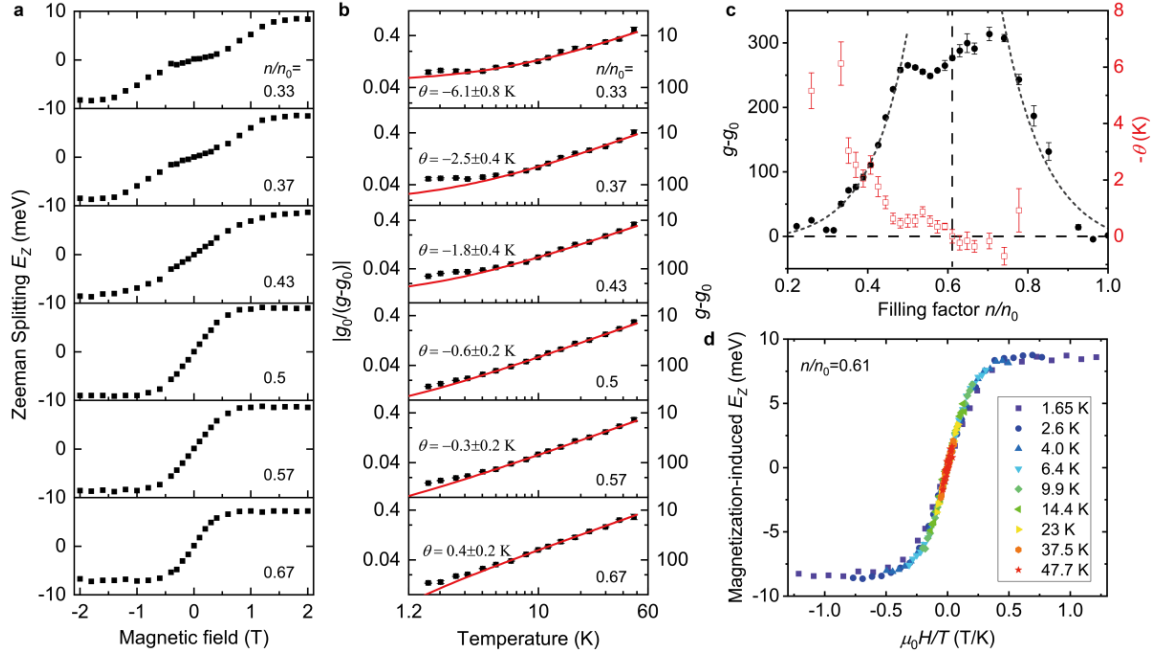

**Figure 4 | Magnetic interactions at varying filling factors. a,** Exciton Zeeman splitting as a function of magnetic field at filling factor ranging from 0.33 to 0.67. **b**, Temperature-dependent $\left|\frac{g_0}{g-g_0}\right|$ in a log-log plot fitted by the Curie-Weiss law (red solid curves) from 6.4 K to 48 K. The red dotted curves in the top three panels are comparisons with the paramagnetic limit of Curie's law. **c**, Filling dependent magnetic susceptibility $(g - g_0) \propto \chi$ (left axis, black filled symbols) and Weiss constant $\theta$ (right axis, red empty symbols). The black dotted lines are a guide to the eye to the divergent susceptibility near 0.6 filling denoted by the vertical dashed line. **d**, Exciton Zeeman splitting (excluding the bare exciton contribution) as a function of temperature-scaled magnetic field at 0.61 filling at varying temperatures. All curves collapse, showing that temperature is the only relevant energy scale.

**Extended data figures**

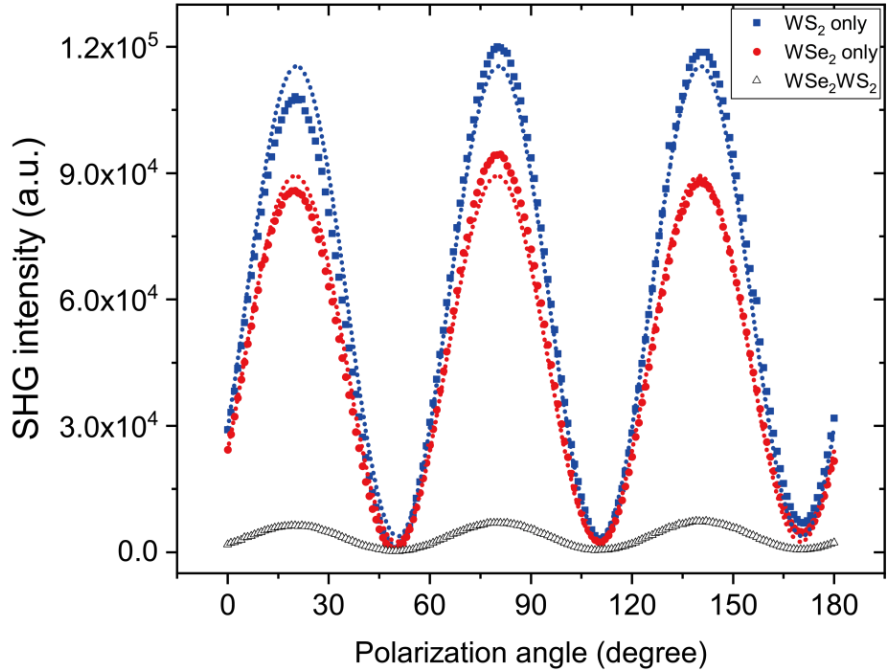

**Extended data figure 1 | Determination of crystal orientations by angle-resolved second harmonic generation (SHG).** Solid squares and circles are polarization dependent SHG from regions of monolayer WS$_2$ and WSe$_2$, respectively. Dotted curves are fits to the data using $C + A cos^2[3(\varphi - \varphi_0)]$, where $\varphi$ is the excitation polarization angle, $\varphi_0$ (initial crystal orientation), $C$ and $A$ are three fitting parameters. The initial crystal orientation $\varphi_0$ is determined to be $20.4° \pm 0.1°$ and $20.0° \pm 0.1°$ for WS$_2$ and WSe$_2$, respectively. The twist angle is therefore $0.4° \pm 0.2°$. Empty triangles are SHG from the overlapped region of the two materials. Since it is much weaker than the SHG from the monolayer-only regions, we conclude that WS$_2$ and WSe$_2$ are nearly 60° aligned (*i.e.* AB stacking), in which case the second-harmonic dipoles from the two layers are out-of-phase.

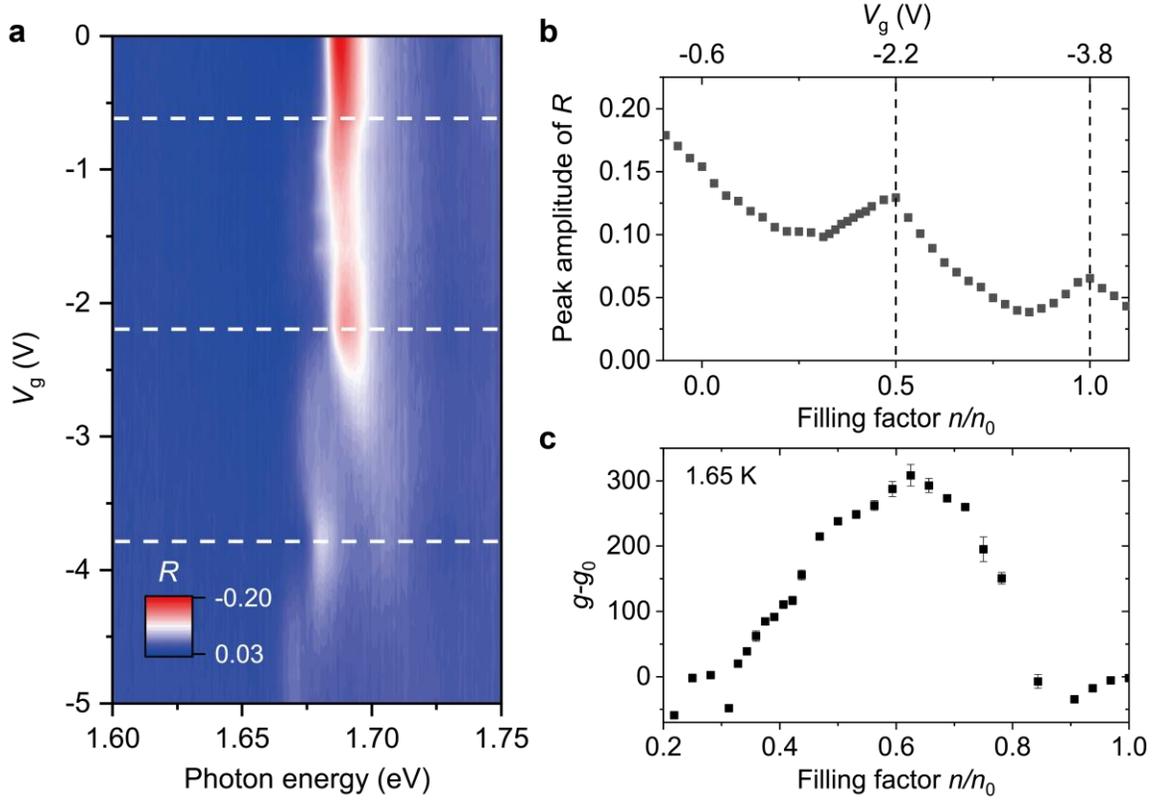

**Extended data figure 2 | Behavior of a 0° aligned WSe$_2$/WS$_2$ bilayer. a**, Contour plot of reflectance contrast spectra as a function of gate voltage. As in the 60° aligned sample data (Fig. 2 of the main text), the dashed lines respectively mark, from top to bottom, zero, half and full filling of the first moiré valence band. **b**, Reflection peak amplitudes at the lowest-energy exciton resonance as a function of filling at 1.65 K. The two peaks, denoted by two dashed lines, occur at half and full filling. **c**, Filling-dependent g-factor at 1.65 K. The error is from linear fitting of magnetic-field dependent Zeeman splitting.

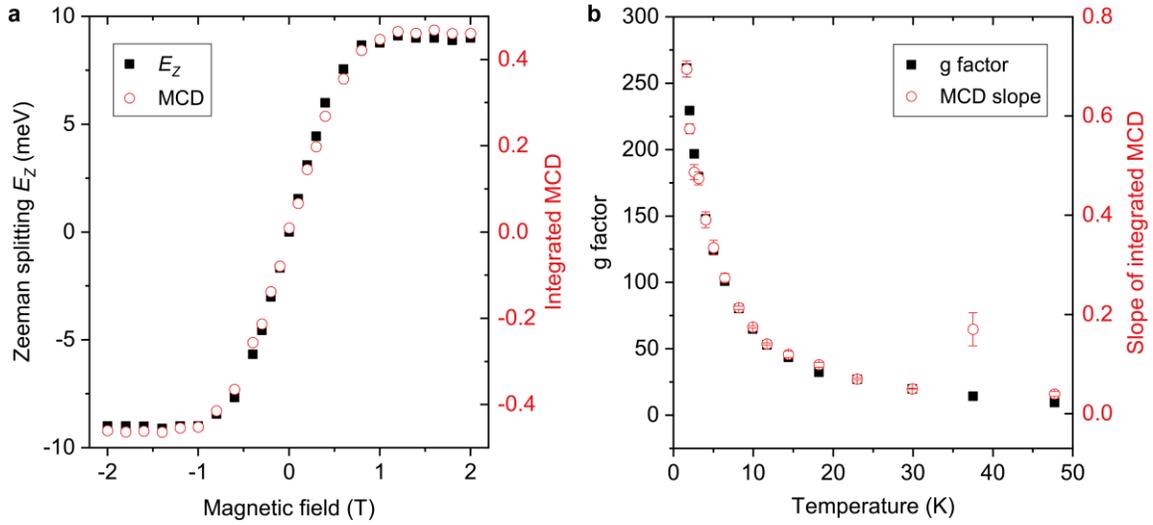

**Extended data figure 3 | Comparison between Zeeman splitting and magnetic circular dichroism (MCD) analysis at half filling.** **a**, Black filled squares and red empty circles denote Zeeman splitting $E_Z$ and spectrally integrated MCD signal respectively as a function of magnetic field. The integrated MCD is defined in the Methods section. An MCD contrast approaching 0.5 has been obtained above 1 T. **b**, The black filled squares and red empty circles respectively denote the temperature dependence of the g-factor (obtained from $E_Z$) and the slope of the MCD contrast in **a** in the small field limit. Overall, the results from the two different analysis methods agree well with each other.

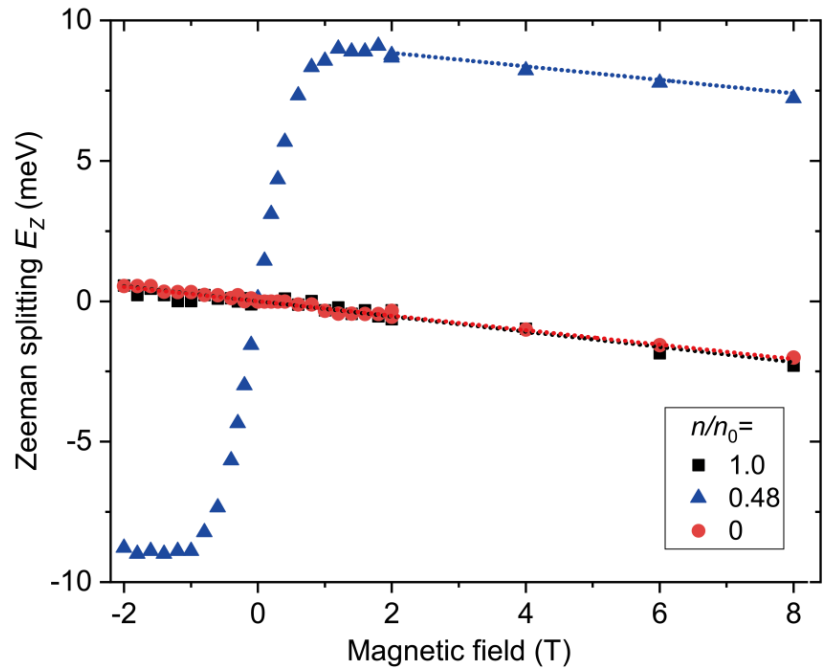

**Extended data figure 4 | Zeeman splitting as a function of magnetic field up to 8 T at zero, half, and full fillings.** The black, blue and red symbols are, respectively, Zeeman splitting at full, half and zero fillings. The black, blue and red dotted lines are the corresponding linear fits to Zeeman splitting, revealing g-factor of about -4.7, -4.1 and -4.4. Only the high-field region of the half-filling data is used for the fitting.

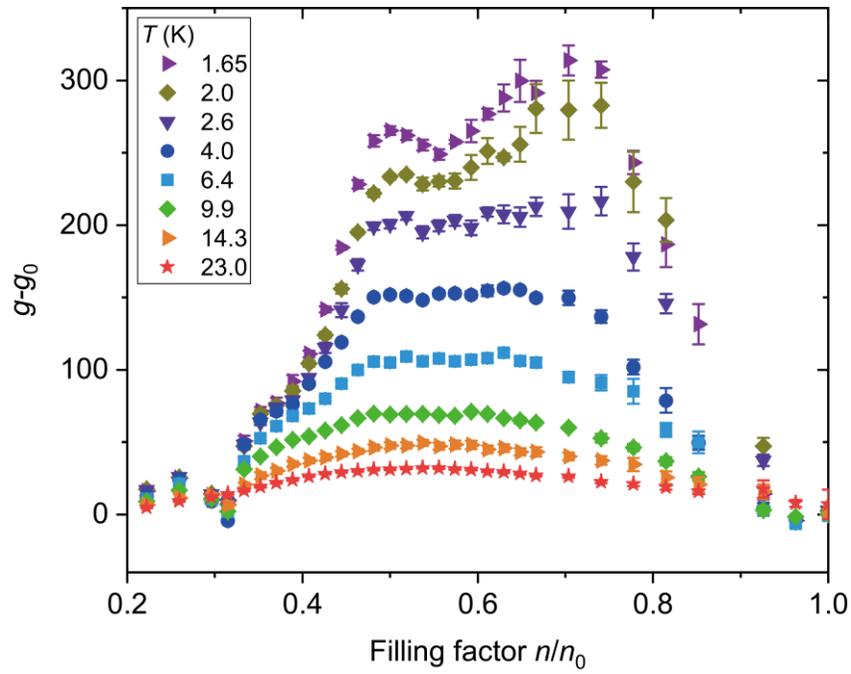

**Extended data figure 5 | Doping-dependent g-factors at varying temperatures.** The errors are from linear fitting of magnetic-field-dependent Zeeman splitting at small magnetic field.

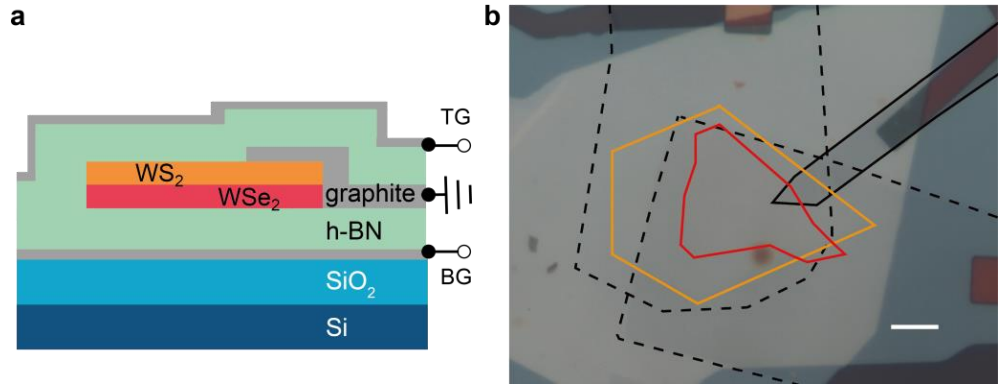

**Extended data figure 6 | Schematic and optical image of a representative device. a**, Schematic device geometry of a dual-gated $WSe_2/WS_2$ bilayer. **b**, Optical image of a typical device on a $SiO_2/Si$ substrate. The red and orange solid lines denote the sample edges of $WSe_2$ and $WS_2$, respectively. The dashed black lines denote the top and bottom graphene gates. The solid black line denotes the graphene contact electrode. The scale bar is 5 μm.